\documentstyle[epsf,12pt]{article}
\begin{document}
\begin{center}
\Large{\bf Observation of $\Delta \phi \Delta \eta$ Scaled Correlation Signals
which increase with Centrality of Au Au collisions at $\sqrt{s_{NN}}$ = 200 GeV
}\\
\large{R.S. Longacre$^a$ for the STAR Collaboration\\
$^a$Brookhaven National Laboratory, Upton, NY 11973, USA\footnote{This research
was supported by the U.S. Department of Energy under Contract No. 
DE-AC02-98CH10886}}
\end{center}
 
\begin{abstract}
We show the preliminary charged-particle pair correlation analyses presented in
a poster session at the 2006 International Quark Matter Conference in Shanghai 
China. The correlation analysis space of $\Delta \phi$ (azimuth) and 
$\Delta \eta$ (pseudo-rapidity) are considered as a function of centrality for 
minimum bias Au + Au collisions in the mid-transverse momentum range in the 
STAR detector. The analyses involve unlike-sign charge pairs and like-sign 
charge pairs, which are transformed into charge-dependent (CD) signals and 
charge-independent (CI) signals. We use a multiplicity scale to compare the 
different centralities. We find the signals increase with increasing 
centrality. A model featuring dense gluonic hot spots as first proposed by van 
Hove predicts that the observables under investigation would have sensitivity 
to such a substructure should it occur. A blast wave model including multiple 
hot spots motivates the selection of transverse momenta in the range 
0.8 GeV/c$ < p_t < $4.0 GeV/c.

\end{abstract}
 
\section{Introduction} 

    The search for a Quark-Gluon Plasma (QGP) has a high priority at the
Relativistic Heavy Ion Collider (RHIC). Van Hove\cite{van} and 
others\cite{bubble} have proposed that bubbles localized in phase space ( dense
gluon-dominated hot spots) could be the sources of final state hadrons from a 
QGP. Such structures would have smaller spatial dimensions than the region of 
the fireball. Correlations resulting from these smaller structures might 
persist in the final state of the collision. The Hanbury-Brown and 
Twiss (HBT) results demonstrate that for $\sqrt{s_{NN}}$ = 200 GeV mid 
rapidity central Au + Au, when $p_t > $ 0.8 GeV/c the average final state space
geometry for pairs close in momentum is approximately described by dimensions 
of around 2 fm\cite{HBT}. This should lead to observable modification of the  
$\Delta \eta \Delta \phi$ correlation.
 
\section{$\Delta \phi$ Correlation for Minimum Bias Data}

The $\Delta \phi$ correlation function\cite{KKG} is defined as:

\begin{equation}
C(\Delta \phi)=S(\Delta \phi)/M(\Delta \phi).
\end{equation}
 
Where S($\Delta \phi$) is the number of pairs at the corresponding
values of $\Delta \phi$ coming from the same event, after we have
summed over all the events. M($\Delta \phi$) is the number of pairs
at the corresponding values of $\Delta \phi$ coming from the mixed
events, after we have summed over all our created mixed events. A mixed event 
pair has each of the two particles chosen from a different event. 

\bf Table I. \rm The minimum bias data is split up into standard STAR binning.
\begin{center}
\begin{tabular}{|c|c|}\hline
\multicolumn{2}{|c|}{Table I}\\ \hline
Au + Au Collisions & Centrality \\ \hline
70\% to 80\% & peripheral  \\ \hline
60\% to 70\% &   \\ \hline
50\% to 60\% &   \\ \hline
40\% to 50\% &   \\ \hline
30\% to 40\% &   \\ \hline
20\% to 30\% &   \\ \hline
10\% to 20\% &   \\ \hline
5\% to 10\% &   \\ \hline
0\% to 5\% & most central  \\ \hline
\end{tabular}
\end{center}                                     

The mixed events were selected based on centrality and the Z (beam axis) 
position of the primary vertex. The events selections was achieved by sorting
events into the centrality bins above and ten 5cm wide bins covering -25cm to 
+25cm in Z vertex. Thus only tracks were mixed for events which had the same 
multiplicity and acceptance.
 
\section{The CD and The CI Correlations}

Using equation 1, we can form $\Delta \phi$ correlations for unlike-sign charge
pairs (US) and like-sign charge pairs (LS). We can also form a $\Delta \phi$ 
correlation using all the particle pair combinations which would be charge 
independent. This CI (Charge Independent) correlation is equal to the average
of the US and LS correlations.

\begin{equation}
CI = (US + LS)/2.
\end{equation}
 
We also can form a Charge Dependent (CD) correlation by taking the difference 
between the US and the LS correlations.

\begin{equation}
CD = US - LS.
\end{equation}
 
The CI correlation is a quantitative measure of the average structure of the
correlated emitting sources. The CD correlation is a qualitative measure of the
emission correlation of unlike-sign charge pairs emitted from the same space 
and time region. Simular CI and CD are used by other STAR analyses\cite{CD&CI}.

\section{How does one compare different Centralities?}

The signal present in the correlation comes from pairs that are correlated.
The number of signal pairs increase linearly with the number of particles. The 
total number of pairs is proportional to the square of the number of particles.
Thus the ratio of signal pairs to all pairs decrease as 1/particles. We can 
cancel out this effect by multiplying the CD and CI correlations by the 
multiplicity of particles used in forming the correlations.

\section{Comparing the CD Correlations}
We compare the CD yield in Figure 1 and the overall shape of the CD in Figure 
2. Multiplicity scaling is used in Figure 1 while we normalize each centrality
in Figure 2. In Figure 2 the $\Delta \phi = 10^\circ$ for 0.0 $< \Delta \eta <$
0.3 is normalized to 1. All other $\Delta \eta$ bins are determined by this 
normalization.

\begin{figure}
\begin{center}
\mbox{
   \epsfysize 8.4in
   \epsfbox{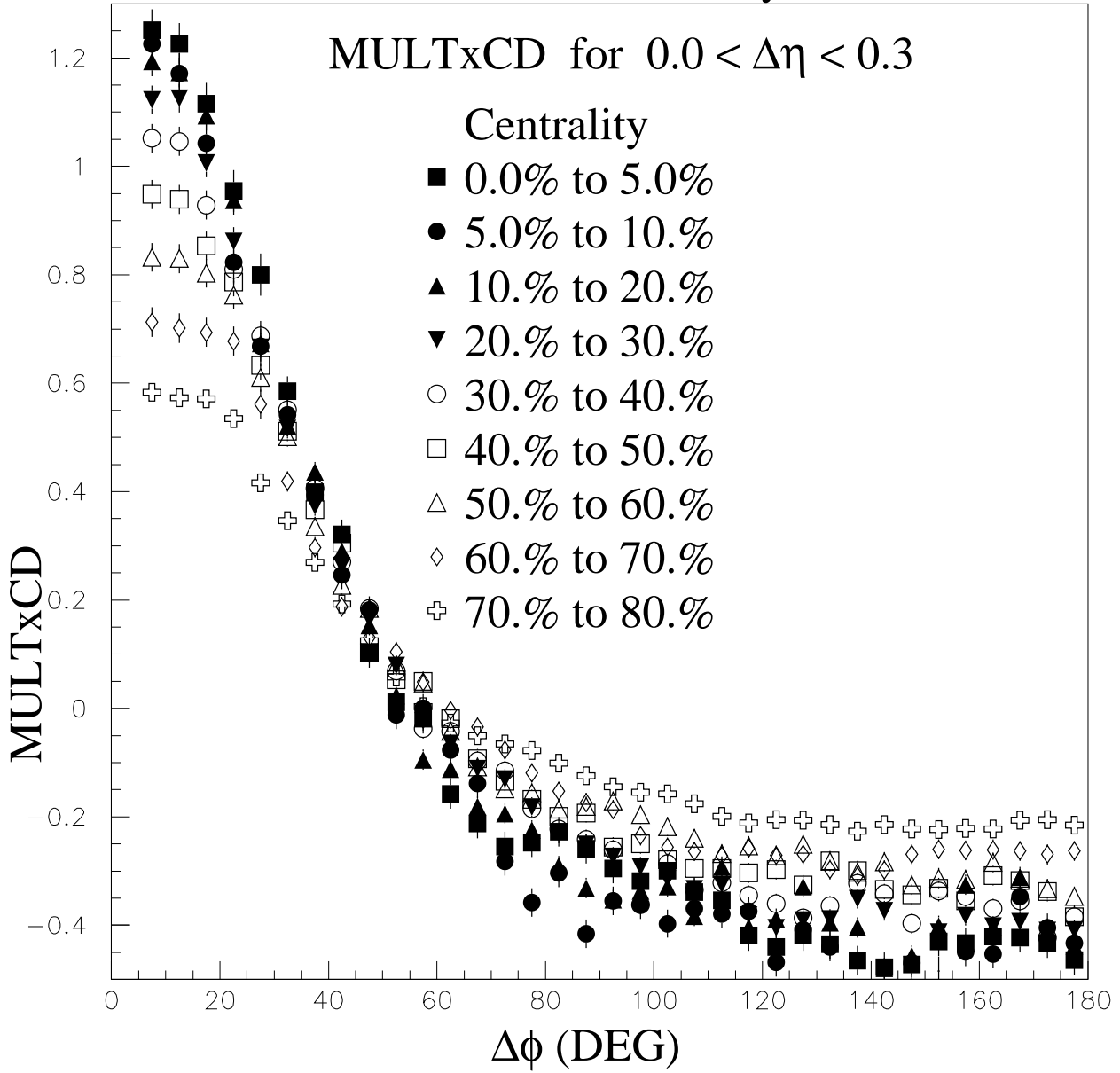}}
\end{center}
\vspace{2pt}
\caption{The multiplicity(MULT) times the CD correlation vs. $\Delta \phi$ for
0.0 $< \Delta \eta <$0.3. Nine centralities are shown from 70\% to 80\% 
increasing to 0\% to 5\%. The CD increases with centrality.}
\label{fig1}
\end{figure}

\begin{figure}
\begin{center}
\mbox{
   \epsfysize 8.4in
   \epsfbox{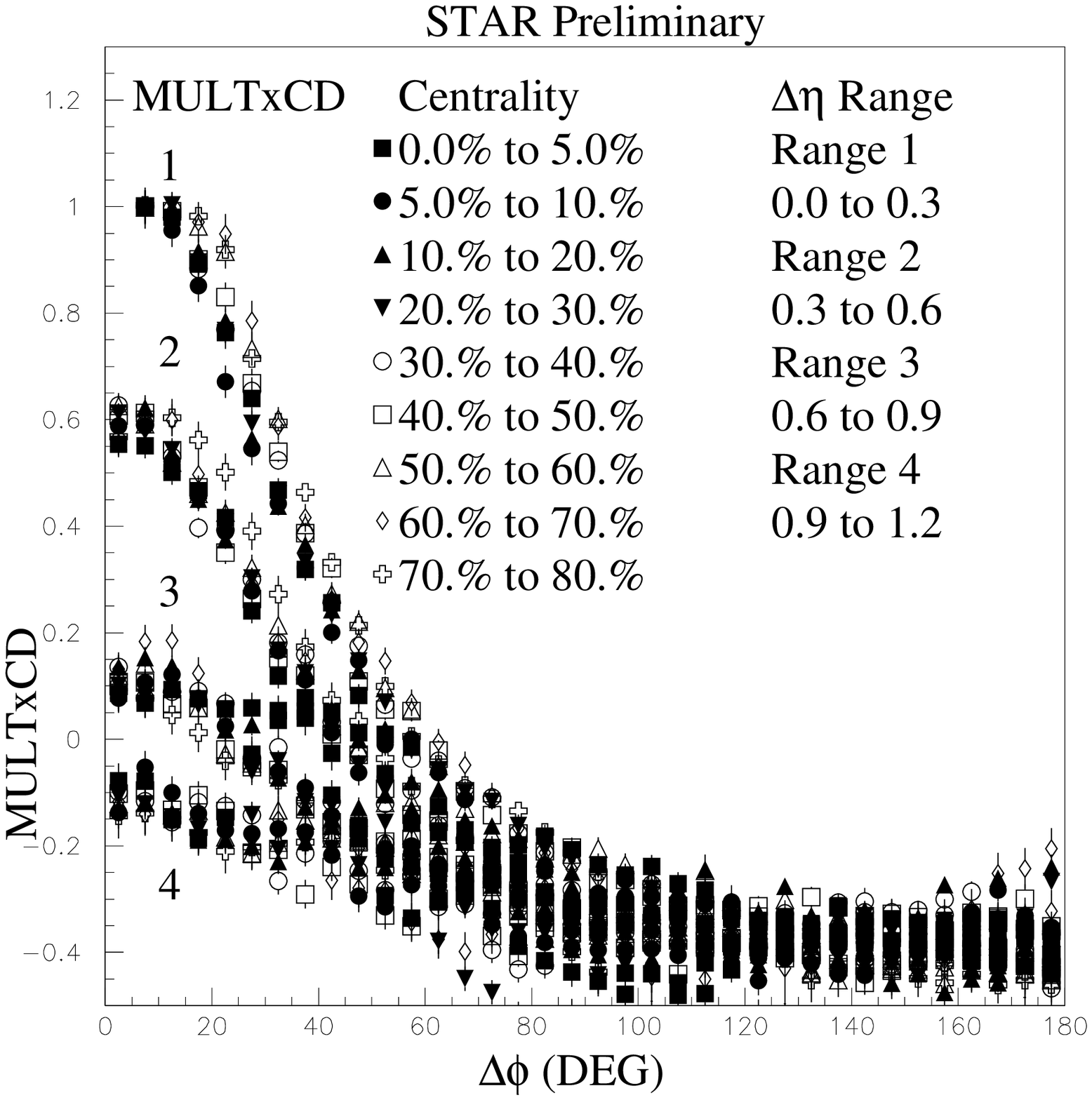}}
\end{center}
\vspace{2pt}
\caption{The multiplicity(MULT) times the CD correlation vs. $\Delta \phi$ for
all $\Delta \eta$ bins.0.3. Nine centralities are shown from 70\% to 80\% 
increasing to 0\% to 5\%. Each of the nine centralities were scaled so that 
the $10^\circ \Delta \phi$ bin for 0.0 $< \Delta \eta <$0.3 is normalized to 1.
We see that the shape of the CD correlation is virtually independent of 
centrality.}
\label{fig2}
\end{figure}
      
\section{Comparing the CI Correlations}

We need to compare the CI yield for the different centralities. The CI 
correlation as a function of $\Delta \phi$ averages to 1 over the $\Delta \phi$
range. In order to compare the signal strength we must multiply by the 
multiplicity for each centrality. The scale correlation will then have a much 
different range and be hard to compare. The average multiplicity ranges from 6 
to 216. We choose to shift up each scaled correlation until it is in the 216 
range.

\subsection{Elliptic Flow}

Elliptic flow is a background that we need to worry about. The elliptic flow is
given by $2 v_2^2\cos (2 \Delta \phi)$ and is symmetric about $\Delta \phi$ = 
$90^\circ$. For a constant value of elliptic flow ($v_2$), the scaled 
correlation will get a bigger dip at $\Delta \phi$ = $90^\circ$ with increasing
multiplicity. The dip is proportional to the multiplicity.

Elliptic flow ($v_2$) decreases linearly with centrality and the correlation 
has a quadratic dependence on $v_2$. Thus for the scaled correlation the 
elliptic flow effect increases because the multiplicity is increasing and then
decreases because of the quadratic response of $v_2$, becoming very small at 
the most central (0\% to 5\%).

\subsection{Determining the Shift Value}

To first order we want to remove the $v_2$ effect from the comparison. If the
away side of the scaled correlation ($\Delta \phi$ = $180^\circ$) is mainly
$v_2$ then a signal on the near side ($\Delta \phi$ = $0^\circ$) would rise 
above the value of the away side. Thus if we would shift all scaled 
correlations such that the away side is equal, then we could compare the 
signals on the near side.

\subsection{Comparing the CI($\Delta \phi$) vs. $\Delta \eta$}

The scaled CI correlation is split up into 5 $\Delta \eta$ bins (Figure 3 to 
Figure 7). Each centrality for a given $\Delta \eta$ bin is plotted with the 
away side ($\Delta \phi$ = $180^\circ$) scaled correlation shifted to the same 
value. The horizontal line for each centrality shows the shifted average 
multiplicity line which was 1 in the original CI correlation and became equal 
the average multiplicity after becoming the scaled correlation.

We have assumed in our comparison that the elliptic flow accounted for the away
side peak. A bubble model\cite{bubble} predicts that there should be an away 
side correlation between bubbles. Thus our normalization would lead to a 
reduced signal on the near side. The bubble model\cite{bubble} showed that 
there is an away side signal when the width of the away side becomes much wider
then the near side as it does for the most central (0\% to 5\%).  

\section{Conclusions}
 
The scaled CD correlations increase with centrality see Figure 1. The scaled
CI correlations increase until 20\% centrality and remains the same for the 
rest of the centralities. If we assume the wider away side peak is also a 
signal, we have a consistent picture between the scaled CD and CI correlations,
both increasing with centrality. Ref.\cite{bubble} is in good agreement with 
our central CD and CI correlations. The bubble model\cite{bubble} assumes 
localized gluonic hot spots on the surface of the fireball created by the 
central Au + Au collision.

\clearpage 

\begin{figure}
\begin{center}
\mbox{
   \epsfysize 8.4in
   \epsfbox{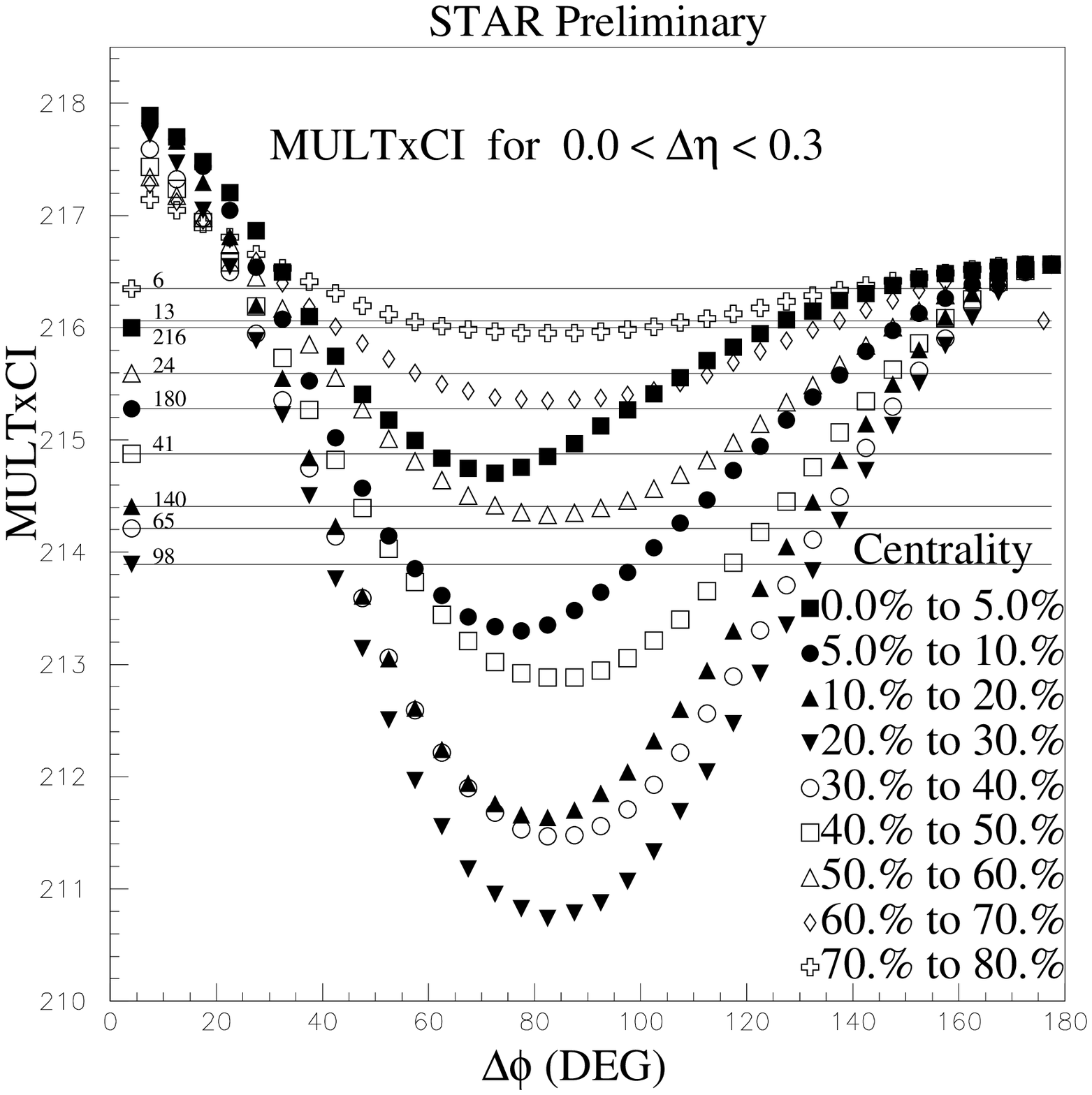}}
\end{center}
\vspace{2pt}
\caption{The multiplicity(MULT) times the CI correlation vs. $\Delta \phi$ for
0.0 $< \Delta \eta <$0.3. Nine centralities are shown from 70\% to 80\% 
increasing to 0\% to 5\%. 216 is the multiplicity for 0\% to 5\% and all other
centralities are shifted up so that the $180^\circ$ value is equal. Each
multiplicity for each centrality is shown shifted.}
\label{fig3}
\end{figure}

\begin{figure}
\begin{center}
\mbox{
   \epsfysize 8.4in
   \epsfbox{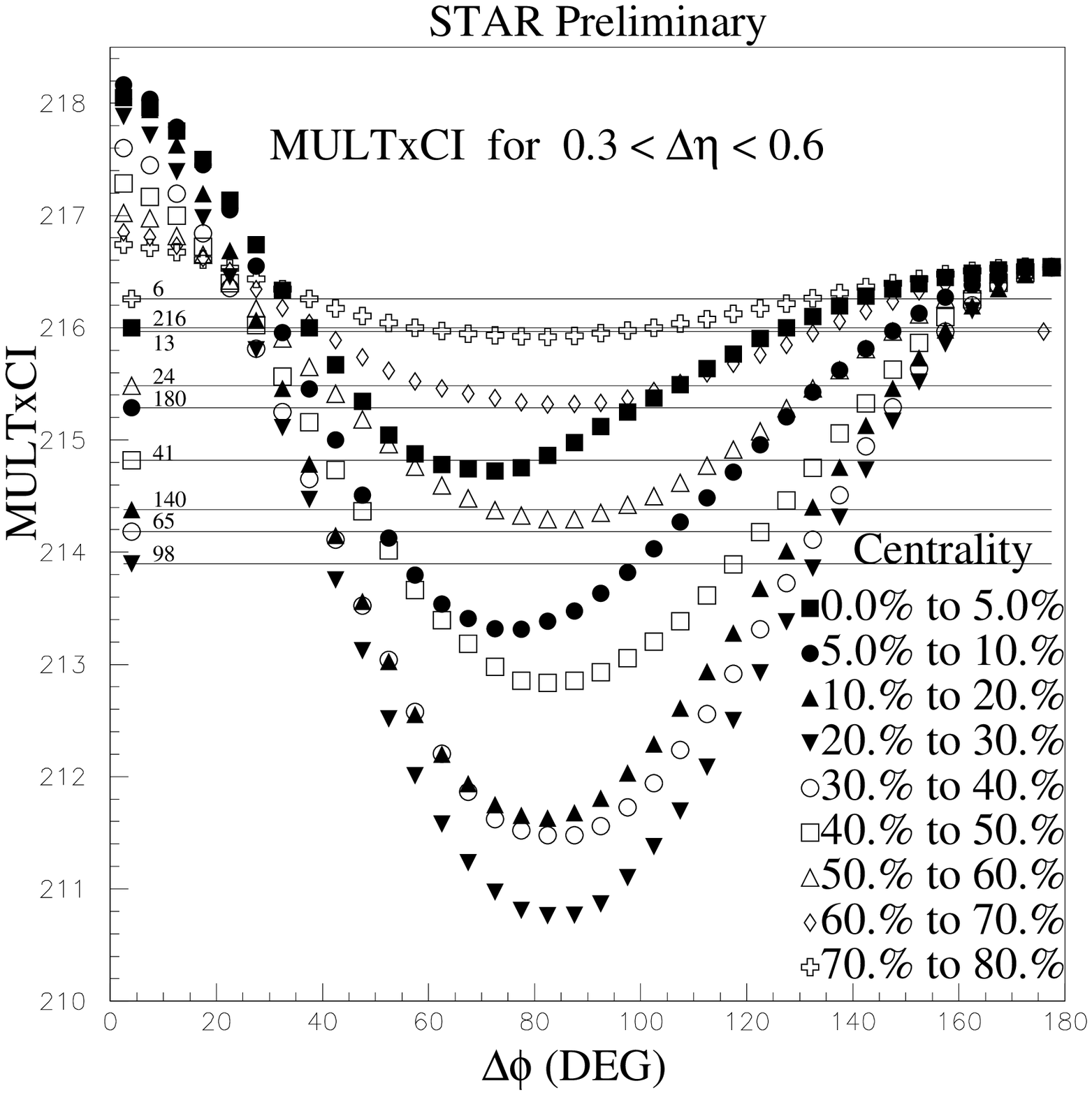}}
\end{center}
\vspace{2pt}
\caption{The multiplicity(MULT) times the CI correlation vs. $\Delta \phi$ for
0.3 $< \Delta \eta <$0.6. Nine centralities are shown from 70\% to 80\% 
increasing to 0\% to 5\%. 216 is the multiplicity for 0\% to 5\% and all other
centralities are shifted up so that the $180^\circ$ value is equal. Each
multiplicity for each centrality is shown shifted.}
\label{fig4}
\end{figure}

\begin{figure}
\begin{center}
\mbox{
   \epsfysize 8.4in
   \epsfbox{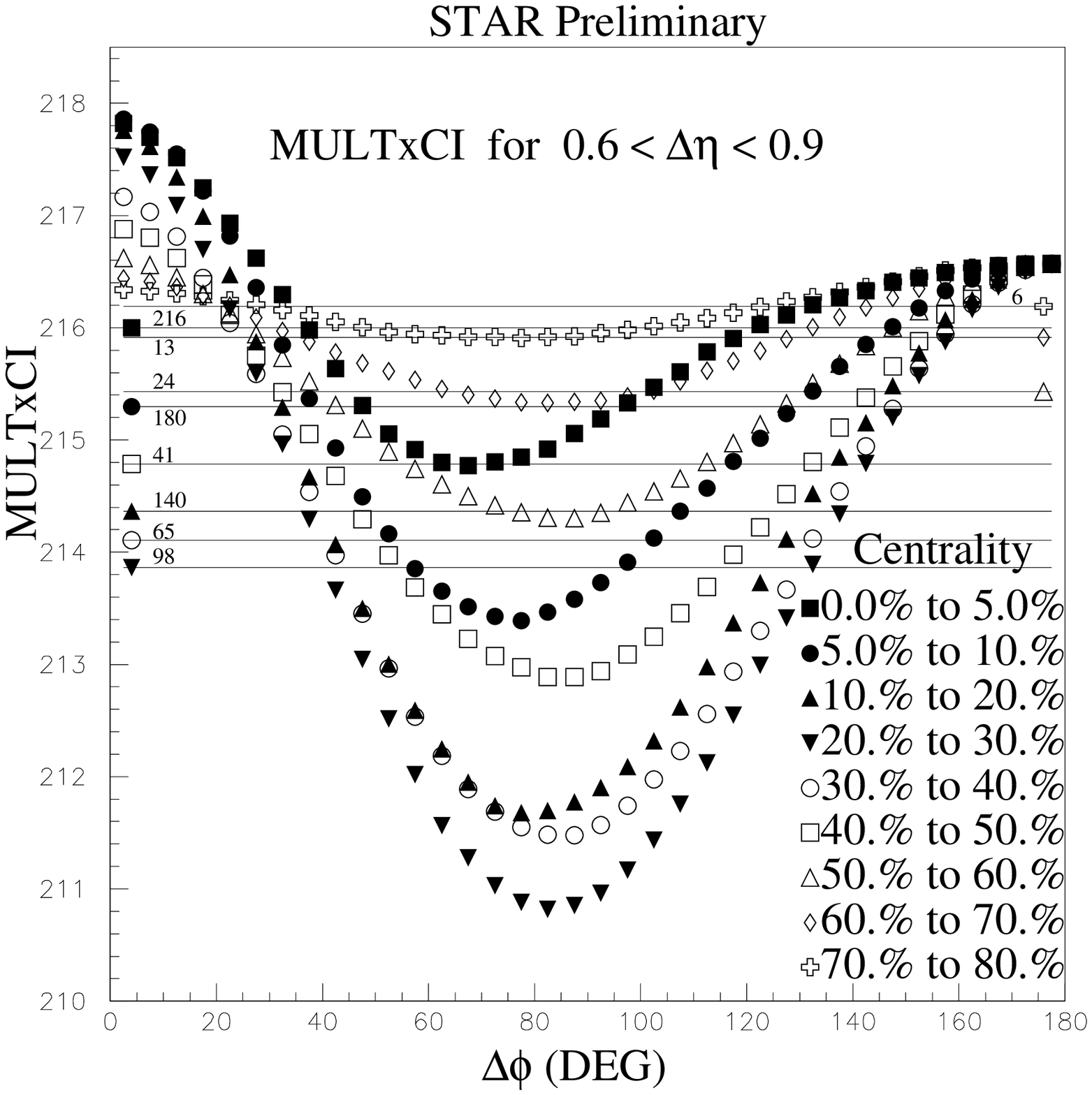}}
\end{center}
\vspace{2pt}
\caption{The multiplicity(MULT) times the CI correlation vs. $\Delta \phi$ for
0.6 $< \Delta \eta <$0.9. Nine centralities are shown from 70\% to 80\% 
increasing to 0\% to 5\%. 216 is the multiplicity for 0\% to 5\% and all other
centralities are shifted up so that the $180^\circ$ value is equal. Each
multiplicity for each centrality is shown shifted.}
\label{fig5}
\end{figure}

\begin{figure}
\begin{center}
\mbox{
   \epsfysize 8.4in
   \epsfbox{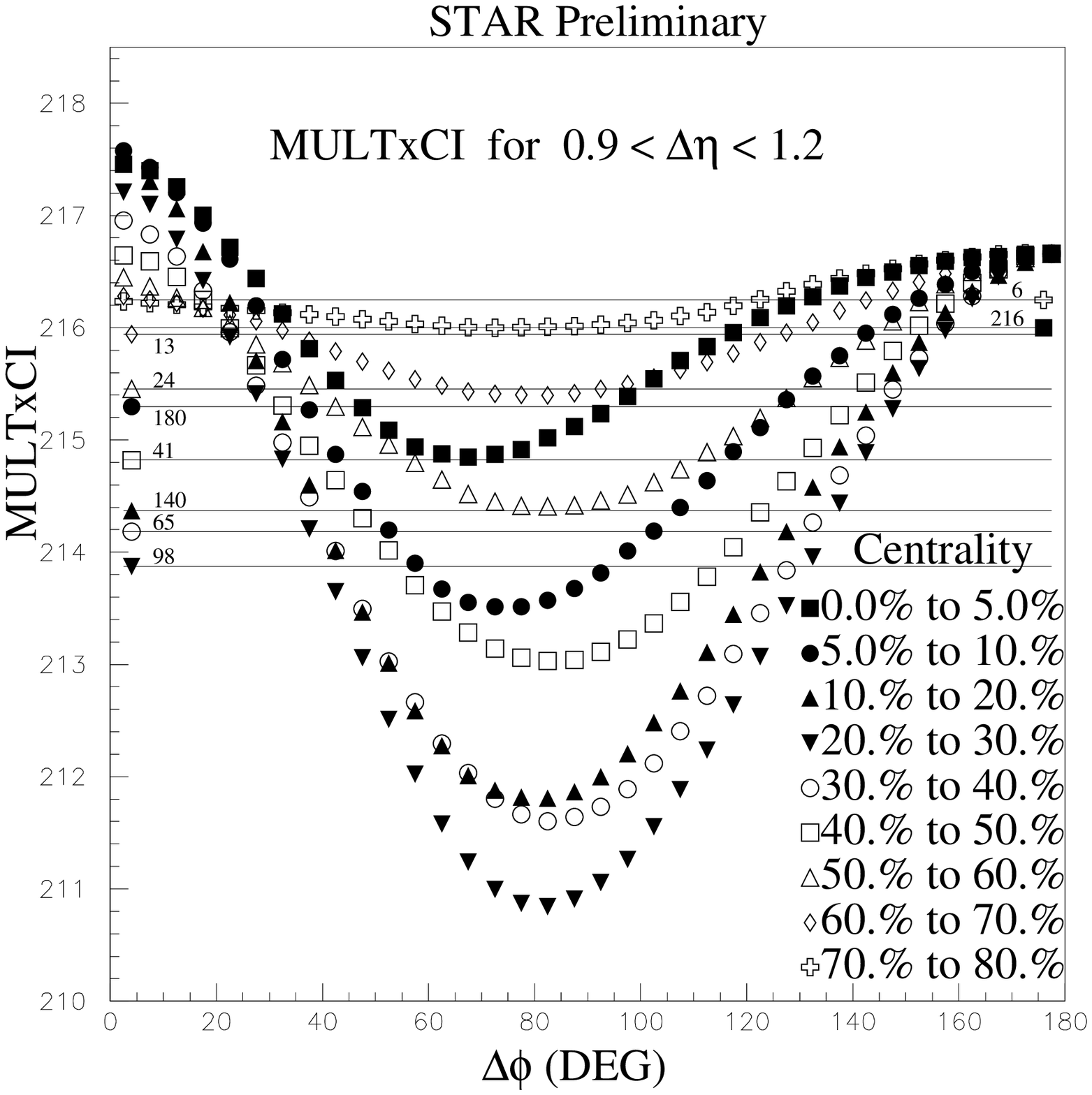}}
\end{center}
\vspace{2pt}
\caption{The multiplicity(MULT) times the CI correlation vs. $\Delta \phi$ for
0.9 $< \Delta \eta <$1.2. Nine centralities are shown from 70\% to 80\% 
increasing to 0\% to 5\%. 216 is the multiplicity for 0\% to 5\% and all other
centralities are shifted up so that the $180^\circ$ value is equal. Each
multiplicity for each centrality is shown shifted.}
\label{fig6}
\end{figure}

\begin{figure}
\begin{center}
\mbox{
   \epsfysize 8.4in
   \epsfbox{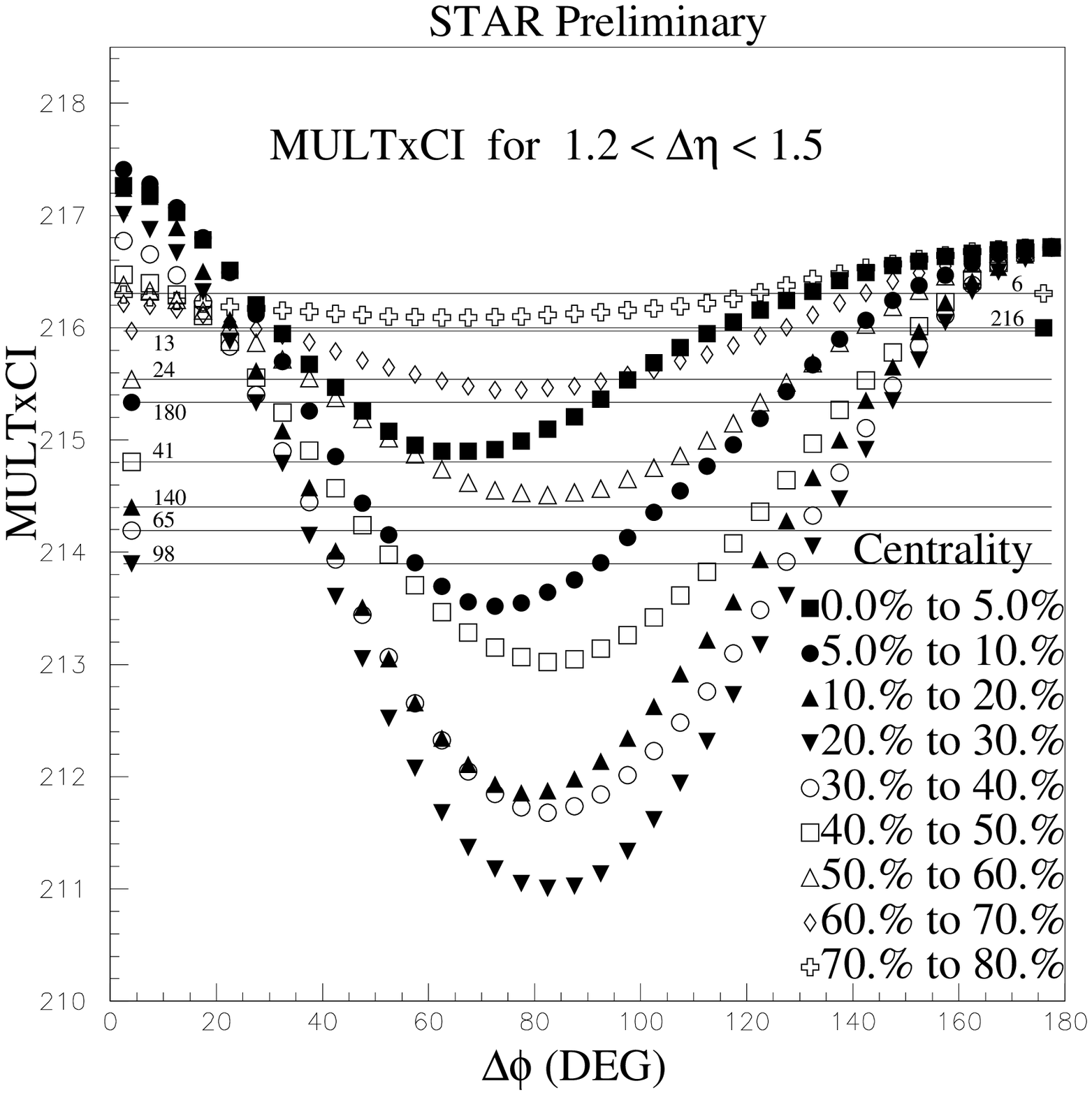}}
\end{center}
\vspace{2pt}
\caption{The multiplicity(MULT) times the CI correlation vs. $\Delta \phi$ for
1.2 $< \Delta \eta <$1.5. Nine centralities are shown from 70\% to 80\% 
increasing to 0\% to 5\%. 216 is the multiplicity for 0\% to 5\% and all other
centralities are shifted up so that the $180^\circ$ value is equal. Each
multiplicity for each centrality is shown shifted.}
\label{fig7}
\end{figure}

\end{document}